\documentclass[10pt, a4paper]{article}
\usepackage[english]{babel}
\usepackage{graphicx}
\usepackage{amssymb}
\usepackage{xcolor}
\usepackage{setspace}
\usepackage[margin=0.75in]{geometry}
\usepackage{amsmath}
\newtheorem{remark}{Remark}
\newtheorem{example}{Example}
\newtheorem{definition}{Definition}

\providecommand{\keywords}[1]
{\textbf{\textit{Keywords:}}#1}
\begin{document}
\doublespacing
\title{Estimation of the Parameters of Multivariate Stable Distributions}
\author{Aastha M. Sathe and N. S. Upadhye\\
\footnotesize{Department of Mathematics, Indian Institute of Technology Madras,
Chennai-600036, INDIA}
}
\date{\today}
\maketitle
\begin{abstract}
\noindent
In this paper, we begin our discussion with some of the well-known methods available in the literature for the estimation of the parameters of a univariate/multivariate stable distribution. Based on the available methods, a new hybrid method is proposed for the estimation of the parameters of a univariate stable distribution. The proposed method is further used for the estimation of the parameters of a strictly multivariate stable distribution. The efficiency, accuracy and simplicity of the new method is shown through Monte-Carlo simulation. Finally, we apply the proposed method to the univariate and bivariate financial data.
\newline
\end{abstract}

\section{Introduction}
\label{sec1}
In finance, economics, statistical physics, and various other engineering fields, we often encounter datasets where the ``fitted" distribution deviates from the normal distribution and exhibits excess skewness, kurtosis and heavy tails. To address this concern, Paul L\'evy \cite{lev}, in his study on Generalized Central Limit Theorem,  introduced a rich class of distributions known as the stable distributions. Each univariate distribution, in this class, is characterized by four parameters, namely $\alpha$,  $\beta$,  $\sigma$, and $\delta$, which, respectively, denote the index of stability, skewness, scale and shift of the distribution. Their respective ranges are given by  $\alpha \in (0,2]$, $\beta \in [-1, 1]$, $\sigma > 0$ and $\delta \in \mathbb{R}$. On the other hand, a $d$-dimensional stable random vector is determined by $\alpha \in (0,2]$, the shift vector $\boldsymbol \delta\in \mathbb{R}^{d}$ and the spectral measure $\Gamma$ (a finite Borel measure) on $S^{d}=\lbrace\mathbf{s}: ||\mathbf{s}||_{2}=1\rbrace$ which denotes a unit sphere in $\mathbb{R}^{d}$.

\vskip 2ex
\noindent
 Next, it is natural to fit these distributions on the datasets showing excess skewness, kurtosis and heavy tails, and this brings us to the problem of estimation of the above mentioned parameters. We now begin our discussion with the introduction to some well-known methods that efficiently estimate the parameters of the univariate stable distribution. This will be followed by the discussion of methods available for the multivariate case.
%
%
\vskip 2ex
\noindent
Fama and Roll \cite{fama} estimate the parameters of symmetric stable distribution (i.e., $\beta =0$) using quantile method which is later generalized and improved by McCulloch \cite{mac} to incorporate the skewed case (i.e., $ \beta \neq 0$). 
Press \cite{press} estimates the parameters using method of moments, while Koutrouvelis \cite{Kout} and Kogon-Williams \cite{kg} estimate the parameters using the characteristic function and regression. DuMouchel \cite{du} proposed the maximum likelihood method which is further studied by Mittnik \textsl{et al.} \cite{mt} and Nolan \cite{jp}.
Though, these techniques are beneficial in modeling heavy-tailed data (see \cite {kat}, \cite{jp}, \cite{nolan}, \cite{jnol}, \cite{wang}) simulation studies reveal certain limitations for each of these methods(see \cite{ad}, \cite{borak}). Mcculloch's quantile method \cite{mac} is computationally faster than the regression-based estimation by Koutrouvelis \cite{Kout} and Kogon-Williams \cite{kg}, but fails to provide an estimate whenever $\alpha <0.6$. The estimates obtained via the method of moments (see Press \cite{press}) are of poor quality and are not recommended for more than preliminary estimation. Koutrouvelis  regression-based method is iterative in nature and requires the use of look-up tables which makes the estimation of the parameters quite complex. Thus, Kogon-Williams \cite{kg} simplify and eliminate the need of numerous iterations and the use of look-up tables thereby making the method considerably faster and better in comparison to the method of Koutrovelis especially near $\alpha =1$ and $\beta \neq 0$. However, the method gives slightly worse estimates for very small $\alpha$. Finally, the maximum likelihood method \cite{du}, \cite{mt}, \cite{jp} seems to give the most accurate estimates but is computationally expensive in comparison to the other methods discussed above.

\noindent
In comparison to the univariate case, not much is known about the estimation of the parameters of the multivariate stable distribution. However, some of the estimation methods specifically focussing on the estimation of the spectral measure are by Rachev and Xin \cite{rach}, Cheng and Rachev \cite{cheng}, Nolan \textsl{et al.} \cite{pan} and Mohammadi \textsl{et al.} \cite{moh}. Their methods are based on the use of characteristic functions. Pivato and Seco \cite{piv}, used spherical harmonic analysis while Teimouri \textsl{et al.} \cite{tei} make use of the $U$-statistic proposed by Fan \cite{fan}. Ogata \cite{ogata} proposed the use of generalized empirical likelihood (GEL) method where they constructed the estimating function by empirical and theoretical characteristic function. 

\noindent
In this paper, for the univariate case, we propose a new hybrid method of estimation which outperforms the above-mentioned methods, and in particular, Kogon-Williams method, both, in terms of accuracy and computational speed. Further, motivated by Nolan \textsl{et al.} \cite{pan}, we use our proposed hybrid method which jointly estimates all the parameters of a strictly multivariate stable distribution and outperforms the methods of Mohammadi \textsl{et al.} \cite{moh} and Teimouri \textsl{et al.} \cite{tei} both in terms of computational efficiency and accuracy of the estimators. The term ``hybrid" is used to reflect the combination and modification in the methods of Press, Koutrouvelis and Kogon-Williams. The efficiency, accuracy and simplicity of this new technique are shown through simulation results. 

\noindent
The paper is organized as follows. In Section \ref{sec2}, we discuss some well-known facts related to multivariate stable distributions. In Section \ref{sec3}, a new hybrid method (univariate case) is proposed which efficiently estimates $\alpha$, $\sigma$ and $\delta$. The estimates found are then used to obtain the estimates of the parameters of a strictly multivariate stable distribution. The estimate of the spectral measure $\Gamma$ is obtained using the empirical characteristic function method for the multivariate case. In Section \ref{sec4}, the new method is compared with some of the well-known methods. Finally, using financial data, the efficiency of the new method,  both for the univariate and multivariate case, is demonstrated in Section \ref{sec5}.
\
\section{Preliminaries and Notations}\label{sec2}
In general, for the univariate stable distributions, closed forms for densities are not available, except for a few well-known distributions \textit{viz}. normal ($\alpha= 2$, $\beta= 0$), Cauchy ($\alpha =1$, $\beta= 0$) and L\'evy ($\alpha= 1/2$, $\beta= 1$). However, closed form representation for the characteristic function of a univariate/multivariate stable distribution is available. We first define the multivariate stable random vector $\mathbf{X}\in\mathbb{R}^{d}$ and the characteristic function representation of the distribution of $\mathbf{X}$. For details, see Samorodnitsky and Taqqu \cite{sam}.
\begin{definition}
A random vector \textbf{X}=$(X_{1},X_{2},X_{3},\cdots,X_{d})$ is said to be a stable random vector in $\mathbb{R}^{d}$ if:$~\forall A>0$ and $B>0$ $\exists~ C>0$ and  $\mathbf{D }\in \mathbb{R}^{d}$ such that :
\begin{equation}
A\mathbf{X}^{(1)}+B\mathbf{X}^{(2)}\stackrel{d}{=} C\textbf{X}+\textbf{D}\label{stabdefM}
\end{equation}
where $\mathbf{X}^{(1)}$ and $\mathbf{X}^{(2)}$ are independent copies of \textbf{X} and $C=(A^{\alpha}+B^{\alpha})^{1/\alpha}$.
\end{definition}
\begin{definition}
A random vector \textbf{X} $\in \mathbb{R}^{d}$ is stable if for any n $\geq$ 2, there is an $\alpha \in$ (0,2] and a vector $\mathbf{D_{n}}$ such that
\begin{equation}
\mathbf{X}^{(1)}+\mathbf{X}^{(2)}+\cdots+\mathbf{X}^{(n)}\stackrel{d}{=}n^{1/\alpha}\mathbf{X}+\mathbf{D}_{n}
\end{equation} \label{stabdef}
where $\mathbf{X}^{(1)},\mathbf{X}^{(2)},\cdots\mathbf{X}^{(n)}$ are independent copies of \textbf{X}.
\end{definition}
The vector \textbf{X} is strictly stable when $\mathbf{D}=\mathbf{0}$ $\forall A>0$ and $B>0$ in (\ref{stabdefM}) and $\mathbf{D}_{n}=\mathbf{0} ~\forall~ n \geq 2$ in (\ref{stabdef}). The vector \textbf{X} is symmetric stable if it is stable and satisfies the relation:
\begin{center}
P$\{$\textbf{X $\in$ A}$\rbrace\stackrel{d}{=} P\{$ \textbf{-X $\in$ A}$\rbrace$
\end{center}
for any Borel set \textbf{A} of $\mathbb{R}^{d}$.
The index $\alpha$ in (\ref{stabdef}) is called the index of stability of the vector $\mathbf{X}$ which represents the tail thickness of the distribution. 

\noindent
We need the following notations to define the characteristic function representation. Let $\phi(\mathbf{t})={\mathbb E}(e^{\iota <\mathbf{X},\mathbf{t}>})$ denote the characteristic function of $\mathbf{X}$, where $\iota$ is the unit imaginary number and $\mathbf{t}\in \mathbb{R}^{d}$. Also, let $\Im(\cdot)$ and $\Re(\cdot)$ respectively denote the imaginary and real part of the argument and 
${\rm sign} (\cdot)$ denote the sign function.
\noindent
The standard parametrization of the characteristic function of a stable random vector $\mathbf{X}$ when $\alpha \in (0,2]$ is as follows
$$\phi(\mathbf{t})={\mathbb E}(e^{\iota <\mathbf{X},\mathbf{t}>})=e^{-\mathbf{I}(\mathbf{t})}, \mathbf{t}\in \mathbb{R}^{d}$$ 
where $ <\cdot~, \cdot>$ denotes the dot product between the two vectors and 
\begin{eqnarray}
{I}(\mathbf{t})=\int_{S^{d}}\psi_{\alpha}(<\mathbf{t},\mathbf{s}>)\Gamma(d\mathbf{s})+\iota<\boldsymbol\delta, \mathbf{t}>
\end{eqnarray}
where
\begin{eqnarray}
\psi_{\alpha}(u)&=& \begin{cases}
|u|^{\alpha}(1-\iota ~ \rm sign (\it u)\tan \frac{\pi \alpha}{2}),&
\alpha \neq 1,\\
|u|(1+\iota \frac{2}{\pi} \rm sign (\it u)\ln |\it u|),  & \alpha = 1.
\end{cases}
\end{eqnarray}
\begin{remark}
The standard parametrization is discontinuous at $\alpha =1$, since $|\tan \pi \alpha/2|\rightarrow 1$ as $\alpha \rightarrow 1$. As a result, $\Gamma$ and $\boldsymbol\delta$ are poorly estimated whenever $\alpha \rightarrow 1$. To overcome this problem, one can use the multivariate version of parametrization given by Zolotarev \cite{zol} also termed as the continuous parametrization defined in Nolan \cite{lan} where
\begin{eqnarray}
\psi_{\alpha}(u)&=& \begin{cases}
| u |^{\alpha}\Big(1+\iota ~ \rm sign ( \it u)\tan \frac{\pi \alpha}{2}(|\it u |^ {1-\alpha}-1)\Big),&
\alpha \neq 1,\\
|u|\big(1+\iota \frac{2}{\pi} \rm sign (\it u)\ln |\it u|\big),  & \alpha = 1.
\end{cases} 
\end{eqnarray}
\end{remark}
\begin{remark} 
The univariate stable random variable $X$ is described by four parameters $(\alpha, \beta, \sigma, \delta)$. The two main characteristic function representations for random variable $X$ are given by 
\begin{eqnarray}
\phi(t)&=& \begin{cases}
\exp\left\{-(\sigma|t|)^\alpha\left[1+\iota \beta{\rm sign}(t)\tan\left(\frac{\pi\alpha}{2}\right)\left((\sigma |t|) ^{1-\alpha}-1\right)\right]+ \iota \delta \textit{t}\right\},&
\alpha \neq 1,\\
\exp\left\{-\sigma|t|\left[1+\iota \beta \frac{2}{\pi}{\rm sign}(t)\ln(\sigma|t|)\right]+ \iota \delta t \right\},  & \alpha = 1;
\end{cases} \label{0par}\\
\phi( \textit{t})&=& \begin{cases}
\exp\left\{-(\sigma|t|)^\alpha\left[ 1 - \iota \beta{\rm sign}(t)\tan\left(\frac{\pi\alpha}{2}\right)\right]+\iota \delta_{1} t\right\}, &~~~~~~~~~~~~~~~~~~~\alpha \neq 1,\\
\exp\left\{-\sigma|t|\left[ 1+ \iota \beta \frac{2}{\pi}{\rm sign}(t)\ln|t|\right]+ \iota \delta_{1} \textit{t} \right\},  &~~~~~~~~~~~~~~~~~~~\alpha=1,
\end{cases}\label{1par}
\end{eqnarray}
where $
 \delta_{1}=\begin{cases}
~~\delta+\beta\sigma\tan\frac{\pi\alpha}{2},&\hspace{0.2cm}\alpha \neq 1\\
~~\delta+\beta\frac{2}{\pi}\sigma \ln \sigma, &\hspace{0.2cm}\alpha =1.
\end{cases}
$
\newline
Nolan \cite{jnol} recommends (\ref{0par}) for numerical computations and statistical analysis, as the characteristic function is jointly continuous in all parameters, while  (\ref{1par}) can be used in the study of theoretical properties of the stable distribution.
\end{remark}
\section{Estimation of the Parameters of the Strictly Stable Random Vector}\label{sec3}
Our main goal is  to efficiently and accurately estimate the spectral measure $\Gamma$, the shift vector $\boldsymbol\delta$ and the characteristic exponent $\alpha$, given $\mathbf{X}^{(1)},\mathbf{X}^{(2)},\cdots\mathbf{X}^{(n)}$ as independent copies of $\textbf{X}$, a $d$-dimensional stable random vector. Throughout this paper, assume $\boldsymbol\delta=\mathbf{0}$ by replacing $\mathbf{X}$ with $\mathbf{X}-\boldsymbol\delta$.

\subsection{Estimation of $\alpha$ and $\boldsymbol{\delta}$}
For the estimation of $\alpha$ and  the shift vector $\boldsymbol \delta$, Nolan \textsl{et al.} \cite{pan} suggested using some method to estimate the one-dimensional parameters $(\hat{\alpha}_{j}, \hat{\beta}_{j}, \hat{\sigma}_{j}, \hat{\delta}_{j})$, $j=1, 2, \cdots, d$ for each of the coordinates of the $d$-dimensional dataset. In \cite{pan}, the vector $\hat{\boldsymbol\delta}= (\hat{\delta}_{1},\hat{\delta}_{2}, \cdots, \hat{\delta}_{d})$ is used as an  estimate of the shift vector and $\hat{\alpha}=(\sum_{j=1}^{d}\hat{\alpha}_{j})/d$ is used as an estimate of the joint index of stability $\alpha$. 
We now discuss our proposed hybrid method (univariate case) which efficiently estimates the tail index $\alpha$ and the shift vector $\boldsymbol \delta$.
\subsubsection{Proposed Hybrid Method-Univariate Case}
\begin{itemize}
\item[\textbf{Step 1.}] Given a sample of iid observations $x_{1}, x_{2}, \cdots, x_{n}$, obtain the initial estimates $\hat{\alpha}_{0}, \hat{\sigma}_{0}, \hat{\delta}_{0}$ of  $\alpha$, $\sigma$ and $\delta$ respectively using the method of Kogon-Williams which makes use of the continuous parametrization of the characteristic function as defined in (\ref{0par}). Normalize the sample data with the initial estimates of scale ($\hat{\sigma}_{0}$) and shift parameter ($\hat{\delta}_{0}$)
$$x_{j}' = \frac{x_{j}-\hat{\delta}_{0}}{\hat{\sigma}_{0}}, j=1, 2, \cdots, n $$
The above normalization is necessary, in order to remove the dependence of the estimators on $\sigma$ and $\delta$ as originally suggested by Paulson \textsl{et al.} \cite{Paul} and for the optimal selection of the sample characteristic arguments.
\item[\textbf{Step 2.}] Compute the sample characteristic function $\hat{\phi}(t)=\frac{1}{n}\sum_{j=1}^{n}e^{\iota t x_{j}'}$ of the normalized sample $x_{1}', x_{2}',...., x_{n}'$. From (\ref{1par}), observe that, for $\alpha \neq 1$
\begin{equation}
\ln(-\ln|\phi(t)|^2)=\ln(2\sigma^{\alpha})+\alpha \ln|t| \label{eq1}
\end{equation}
Using (\ref{eq1}), obtain the estimates of $\alpha$ and $\sigma$ for the normalized sample data using ordinary least squares regression in the model
\begin{equation}
y_{k}=\mu+\alpha a_{k}+\epsilon_{k},~k=1, 2, \cdots, K,\label{eq2}
\end{equation}
where $y_{k}=\ln(-\ln |\hat{\phi}(t_{k})|^2)$, $\mu = \ln(2\sigma^{\alpha})$, $a_{k}=\ln |t_{k}|$, $\epsilon_{k}$ denotes the error term and $K$, $t_{k}$ are points chosen according to the look-up Table \ref{1}.
\begin{table}[h]
\centering
\begin{tabular}{ | c | c | c | c |}
 \hline
$\alpha$ & $n=200$ & $n=800$ & $n=1600$ \\
\hline
1.9  & 9    &9  &10\\
1.5 & 11 &11&11\\
1.3 & 22 & 16 &14  \\
1.1  &   24      &18    &15\\
0.9  &   28     &22     &18\\
0.7 &   30    &24     &20\\
0.5  &   86      &68    &56\\
0.3 &   134    &124     &118\\
\hline
\end{tabular} 
\caption{Optimum number $K$ of points, $t_{k}= \pi k/25$, $k=1, 2, \cdots, K$}
\label{1}
\end{table}
Let $\hat{\alpha}_{1}$ and $\hat{\sigma}_{1}$ denote the regression estimates of $\alpha$ and $\sigma$ respectively. For finding the optimal value $K$, we make use of  Koutrouvelis look-up Table \ref{1} which relates the values of the sample characteristic function argument to the values of the index of stability $\alpha$ and the sample size. He proposed to use $t_{k}=\frac{\pi k}{25}$, $k=1,~2,~ \cdots,~ K$ for estimating the parameters $\alpha$ and $\sigma$ with $K$ ranging between 9 to 134 for different estimates of $\alpha$ and sample sizes. However, we have modified the procedure of finding $K$ while using the hybrid method to find the estimates. The reason being when we incorporated the original approach of finding $K$, as suggested by Koutrouvelis, in our method, we obtained less accurate estimates.

\noindent 
Obtain three continuous functions/curves corresponding to $\alpha$ and each sample size (\textit{n}=200, 800 and 1600) via the method of least squares regression that best fits the dataset given in Table \ref{1}. The three functions obtained are
$$ f_{1}(\alpha)=24.36\alpha^{-1.47}, f_{2}(\alpha)=20.58\alpha^{-1.43}, f_{3}(\alpha)=122.9\alpha^{4}-648.2\alpha^{3}+1245\alpha^{2}-1040\alpha+335.2 $$
To test which three functions fit the given dataset nicely compute the $R^{2}$ statistic for various functions like a polynomial of degree 1 or more, power and exponential. 
After experimenting with various types of functions, we found that the power function seems to fit the second and third column of Table \ref{1} corresponding to $n=200$ and $n=800$ nicely while the fourth column corresponding to $n=1600$ is best fitted by a polynomial function of degree 4. Since the first two curves are power curves, so log-linear transformations have been made. Using the log-linear form, linear regression on the points given in Table \ref{1} is implemented. The $R^{2}$ values for the two power curves $f_{1}$ and $f_{2}$ are 0.949 and 0.963 respectively, while the third curve $f_{3}$ has  $R^{2}$ value 0.996. The functions are then evaluated at $\hat{\alpha}_{0}$. To find the value of $K$ for intermediate values of $n$(sample size up to 1600) and the functional values, linear interpolation is implemented. For $n>1600$, apply linear extrapolation or subdivide the sample into several groups of size not exceeding 1600 and apply the hybrid method to each of the groups.\\ 

\item[\textbf{Step 3.}] The updated estimate of $\delta$, say $\hat{\delta}_{1}$, is found using the method of Press \cite{press}. For estimation of $\delta$, the sample characteristic function of the normalized sample is evaluated at two points say, $t_{1}$ and $t_{2}$ (both positive and unequal) along with the updated estimate $\hat{\alpha}_{1}$. The points $t_{1}=(3^{2.3})^{3.7}$ and $t_{2}=(3^{2.1})^{3.7}$ were obtained through empirical search as suggested by Krutto \cite{krout} which significantly reduced the mean squared error (MSE) at the time of simulation study.

\noindent
Let $u(t)$ denote the imaginary part of the logarithm of the characteristic function in (\ref{1par}). For $\alpha \neq 1,$
$$u(\textit{t})=\Im (\ln \phi(\textit{t}))=\delta_{1}\textit{t}+\sigma^{\alpha}|\textit{t}|^{\alpha}\beta \rm sign(\textit{t} )\tan\frac{\pi \alpha}{2}=\arctan \Bigg(\frac{\Im(\phi(\textit{t}))}{\Re(\phi(\textit{t}))}\Bigg ).$$
Choose two positive non-zero values $\textit{t}_{1}, \textit{t}_{2}$ such that $\textit{t}_{1}\neq \textit{t}_{2},$
\begin{equation}
\frac{u( \textit{t}_{k})}{\textit{t}_{k}}= \delta_{1} + \sigma^{\alpha} \beta |\textit{t}_{k}|^{\alpha-1} \tan\frac{\pi\alpha}{2},~ k= 1,~2.\label{eq77}
\end{equation}
Solve (\ref{eq77}) for $\delta_{1}$ and  replace $\alpha$ by its estimate $\hat{\alpha}_{1}$ and $u(\textit{t})$ by its sample counterpart to obtain the estimator of $\delta$.\\
Since 
$$\hat{\phi}(\textit{t})=\frac{1}{n}\sum_{j=1}^{n}e^{\iota\textit{t}x_{j}'}= (\frac{1}{n}\sum_{j=1}^{n}\cos\textit{t}x_{j}')+ \iota(\frac{1}{n}\sum_{j=1}^{n}\sin \textit{t}x_{j}'),$$
it follows from the properties of complex numbers that 
$$\tan\hat{\textit{u}}(t)= \frac{\sum_{j=1}^{n}\sin \textit{t} x_{j}'}{\sum_{j=1}^{n}\cos \textit{t} x_{j}'}=\frac{\Im(\hat{\phi}(t))}{\Re(\hat{\phi}(t))}.$$
Thus the final expression of the estimator $\hat{\delta_{1}}$ when $\alpha \neq 1$ is
$$\hat{\delta}_{1}= \frac{|\textit{t}_{2}|^{\hat{\alpha}_{1}-1}\frac{\hat{\textit{u}}(\textit{t}_{1})}{\textit{t}_{1}}-|\textit{t}_{1}|^{\hat{\alpha}_{1}-1}\frac{\hat{\textit{u}}(\textit{t}_{2})}{\textit{t}_{2}}}{(|\textit{t}_{2}|^{\hat{\alpha}_{1}-1}-|\textit{t}_{1}|^{\hat{\alpha}_{1}-1})}.$$
When $\alpha =1$ the estimators have the form 
$$\hat{\delta}_{1}= \frac{\ln|\textit{t}_{2}|\frac{\hat{u}(\textit{t}_{1})}{\textit{t}_{1}}-\ln|\textit{t}_{1}|\frac{\hat{u}(\textit{t}_{2})}{\textit{t}_{2}}}{\ln|\textit{t}_{2}|-\ln|\textit{t}_{1}|}$$
\item[ \textbf{Step 4.}] Compute the final estimates of the sample data as
\begin{center}
$\hat{\alpha}=\hat{\alpha}_{1},~ \hat{\sigma}$=$\hat{\sigma}_{0}\hat{\sigma}_{1},~ \hat{\delta} = \hat{\sigma}_{0}\hat{\delta}_{1}+\hat{\delta}_{0}$
\end{center}
\end{itemize}
\subsection{Estimation of $\Gamma$}
The estimation of the spectral measure is vital in the modeling of stochastic processes. For example, in   portfolio optimization, the dependence structure between the individual stocks is studied and analysed through the spectral measure estimation. More applications can be seen in Tsakalides and Nikios \cite{tk}.\\
Nolan \textsl{et al.} \cite{pan} suggested two methods namely, empirical characteristic function and the projection method for the estimation of the spectral measure. In our proposed method we make use of the empirical characteristic method to get the estimate of $\Gamma$.

\subsubsection{Empirical Characteristic Function Method-ECF}
Given an iid sample $\mathbf{X}^{(1)},\mathbf{X}^{(2)},\cdots\mathbf{X}^{(n)}$ of stable random vectors with the spectral measure $\Gamma$, let $\hat{\phi}_{n}(\mathbf{t})$ and $\hat{I}_{n}(\mathbf{t})$ be the empirical counterparts of $\phi$ and $I$ respectively defined as
$$\hat{\phi}_{n}(\mathbf{t})=(1/n)\sum_{i=1}^{n}e^{\iota< \mathbf{t}, \mathbf{X}^{(i)}>)},~~~ \hat{I}_{n}(\mathbf{t})=-\ln \hat{\phi}_{n}(\mathbf{t})$$
For the estimation of the spectral measure $\Gamma$, Nolan \textsl{et al.} \cite{pan} considered a discrete approximation to the exact spectral measure (see Byczkowski \textsl{et al.}\cite{by}) of the form  
\begin{eqnarray}
\Gamma^{*}=\sum_{l=1}^{L}\gamma_{l}\delta_{\mathbf{s}_{l}} \label{we}
\end{eqnarray}
where $\gamma_{l}=\Gamma(A_{l}), l=1, \cdots, L$ are the weights at point $\mathbf{s}_{l}\in S^{d}$, a unit sphere and $\delta_{\mathbf{s}_{l}}$ is a point mass at $\mathbf{s}_{l}$. The patches that partition the sphere $S^{d}$, with some ``center" $\mathbf{s}_{l}$ are represented by $A_{l}$. Thus, the characteristic function $\phi(\mathbf{t})$ is transformed to $\phi(\mathbf{t})=e^{-\sum_{i=1}^{L}\psi_{\alpha}(<\mathbf{t},~\mathbf{s}_{i}>)\gamma_{i}}$.

\noindent 
Next, for given frequencies $t_{1},\cdots,t_{L} \in \mathbb{R}^{d}$, define an $L \times L$ matrix $\psi$ whose $(k,l)$-th element is $\psi_{\alpha}(<\mathbf{t}_{k},\mathbf{s}_{l}>)$. Finally obtain the expression $\mathbf{I}=\psi \boldsymbol\gamma$, where $\boldsymbol\gamma=(\gamma_{1},\cdots, \gamma_{L})^{'}$. Replacing $\mathbf{I}$ by $\hat{\mathbf{I}}=(\hat{I}(t_{1}),\cdots,\hat{I}(t_{L})$ and choosing $t_{1},t_{2},\cdots,t_{L}$ in such a way that $\psi^{-1}$ exists, we obtain the discretized estimator $\hat{\boldsymbol\gamma}=\psi_{\hat{\alpha}}^{-1}\hat{\mathbf{I}}$ of the spectral measure $\Gamma$.

\noindent
For a general spectral measure $\Gamma$ (not discrete and/or the location of the point masses are unknown)
consider the discrete approximation defined above in (\ref{we}).

\subsubsection{Modifications in ECF method for the estimation of $\Gamma$}
\begin{itemize}
\item[1.] For $d=1$, the location of point masses are concentrated at just two points 1 and -1. We take $\mathbf{s}_{l}=(-1)^{l}$ and $\mathbf{t}_{l}=(-1)^{l+1}$ for $l=1,~2$. For obtaining the discretized estimator $\hat{\boldsymbol\gamma}$ of the spectral measure $\Gamma$, define
\begin{eqnarray}
\psi_{\alpha}(u)&=& \begin{cases}
|u|^{\alpha}(1-\rm sign (\it u)\tan \frac{\pi \alpha}{2}),&\label{8}
\alpha \neq 1,\\
|u|(1+\frac{2}{\pi} \rm sign (\it u)\ln |\it u|),  & \alpha = 1.
\end{cases} 
\end{eqnarray}
and $\hat{\mathbf{I}}=\Big(\Re\big(\hat{I}(t_{1})\big)+\Im\big(\hat{I}(t_{1})\big),~\Re\big(\hat{I}(t_{2})\big)+\Im\big(\hat{I}(t_{2})\big)\Big)$ as suggested by Mohammadi \textsl{et al.} \cite{moh}, (Theorem 3.2).
\item[2.] When $d=2$, we take $\mathbf{t}_{l}=\mathbf{s}_{l}=\big(\cos(2\pi(l-1)/L,\sin(2\pi(l-1)/L\big)\in S^{d}$, and arcs $A_{l}=\big(2\pi(l-(3/2))/L,2\pi(l-(1/2))/L\big),l=1,\cdots,L$. In order to eliminate the problem of imaginary weights $\gamma_{j}$, the properties of $\psi$ and $\mathbf{I}$ and a symmetric grid is used. 

\noindent
\begin{itemize}
\item[2.1] When $L=2m$, let the grid be given by $\mathbf{t}_{l}=\mathbf{s}_{l}=\big(\cos(2\pi(l-1)/L,\sin(2\pi(l-1)/L\big)$. Observe that, $\mathbf{I}_{l}=\bar{\mathbf{I}}_{l+m}$ and the entries of $\psi$ satisfy $\psi_{k,l}=\psi(<t_{k},t_{l}>)=\psi\big(\cos(2(k-l)\pi/L)\big)=\bar{\psi}_{k+m,l}$. Thus, for $l=1,\cdots,m$, $\Re I_{l}=(I_{l}+I_{l+m})/2$ and $\Im I_{l}=-(I_{l}-I_{l+m})/2$. Define the real vector $\boldsymbol{c}=(\Re I_{1},\Re I_{2},\cdots,\Re I_{m},\Im I_{1},\Im I_{2},\cdots,\Im I_{m})^{'}$ and the real $L\times L$ matrix $A=a_{k,l}$ by
\[
a_{k,l}=\begin{cases}
\Re \psi_{k,l},& k=1,\cdots ,m\\
\Im \psi_{k,l},  & k=m+1,\cdots ,L.
\end{cases}
\]
then 
$$\boldsymbol{c}=A\boldsymbol{\gamma}$$
In order to avoid the chance of getting complex or negative values for some of the weights, we use the nnls($\cdot$) library in R that solves the minimization problem
\begin{center}
Minimize $||\boldsymbol{c}-A\boldsymbol{\gamma}||_{2}$
subject to $\boldsymbol{\gamma} \geq 0.$
\end{center}

\noindent
\item[2.2]When $L=2m+1$, again let $\mathbf{t}_{l}=\mathbf{s}_{l}=\big(\cos(2\pi(l-1)/L,\sin(2\pi(l-1)/L\big)$. Then discretized estimator 
$\hat{\boldsymbol\gamma}=\psi_{\hat{\alpha}}^{-1}\hat{\mathbf{I}}$ of the spectral measure $\Gamma$ is $\hat{\boldsymbol\gamma}=|\Re(\psi_{\hat{\alpha}}^{-1}\hat{\mathbf{I}})|$.
\end{itemize}

\item[3.] When $d=3$, the locations of the point masses were expressed in the form 
$$\mathbf{s}_{l}=\Big(\sin(\pi/l)\cos(2\pi(l-1)/L),~\sin(\pi/l)\sin(2\pi(l-1)/L),~\cos(\pi/l)\Big),~l=1,\cdots, L.$$ Here we set $\mathbf{t}_{l}=\mathbf{s}_{l}$ for $l=1,\cdots,L$ and define $\psi_{\alpha}(u)$ as 
\begin{eqnarray}
\psi_{\alpha}(u)=|u|^{\alpha}\label{811}
\end{eqnarray}
where $u=<\mathbf{t}_{k},\mathbf{s}_{l}>$ where $k,l=1, \cdots,L$ and $\hat{\mathbf{I}}=\big(\Re(\hat{I}(t_{1}),~\Re(\hat{I}(t_{2}),\cdots,\Re(\hat{I}(t_{L})\big)$ as suggested by Mohammadi \textsl{et al.} \cite{moh}, (Theorem 3.1). Thus, the discretized estimator $\hat{\boldsymbol\gamma}=\psi_{\hat{\alpha}}^{-1}\hat{\mathbf{I}}$ of the spectral measure $\Gamma$.
\end{itemize}
In a similar fashion, we can  proceed with higher dimensions.

\section{Simulation and Comparative Analysis}
\label{sec4}
\subsection{Performance Analysis of the Proposed Hybrid Method-Univariate Case}
In this section we specifically compare the estimation accuracy of the two methods namely, McCulloch's quantile method (MQ), Kogon-Williams regression method (KR) mentioned above with that of our proposed hybrid method through Monte Carlo simulation. Each method is then applied to a data having a stable distribution. The data is generated by the method of Chambers \textsl{et al.} \cite{cms} and all the simulations have been carried out with \textquotedblleft stabledist\textquotedblright ~and \textquotedblleft StableEstim\textquotedblright ~package in R.\\
For a selected set of values of the parameters $\alpha$, $\beta$, $\sigma$, $\delta$ and the sample size $n$, a simulation is run where 1000 replicates of iid stable random variables each of length $n$ are generated. For each replicate, we then obtain the estimates of the parameters  $\alpha$, $\beta$, $\sigma$ and $\delta$ by implementing various estimation techniques. We have quantitatively  evaluated the performance of the parameter estimators using the mean squared error (MSE) criterion and have calculated the mean and standard deviation (Sd) of the estimates to asses their performance.


\subsubsection*{Estimation of $\alpha$, $\sigma$ and $\delta$}
\begin{figure}
\includegraphics[scale=0.5, angle=0]{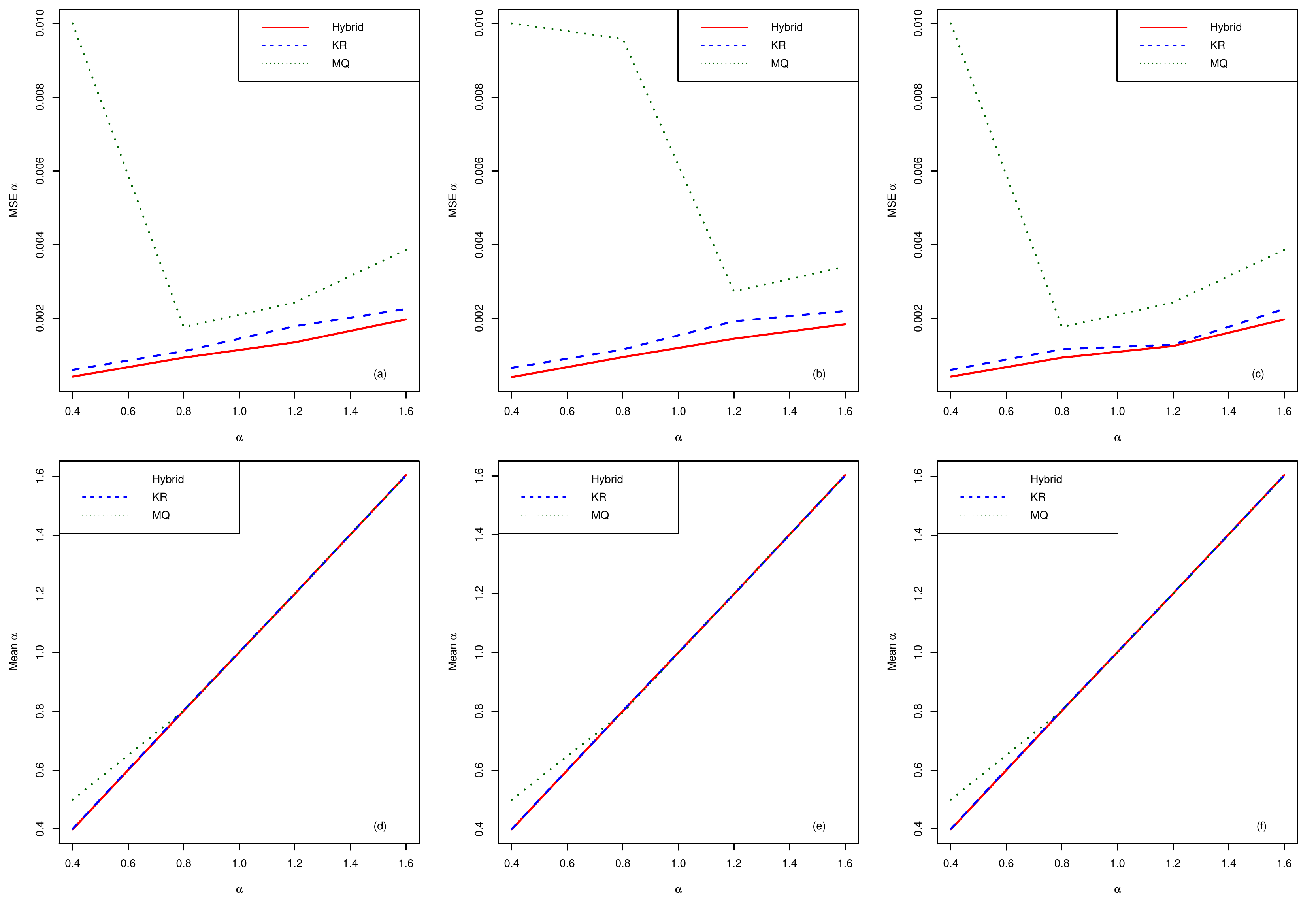}
\caption{Mean and MSE of $\alpha$}
\label{111}
\end{figure}

\begin{figure}
\includegraphics[scale=0.5, angle=0]{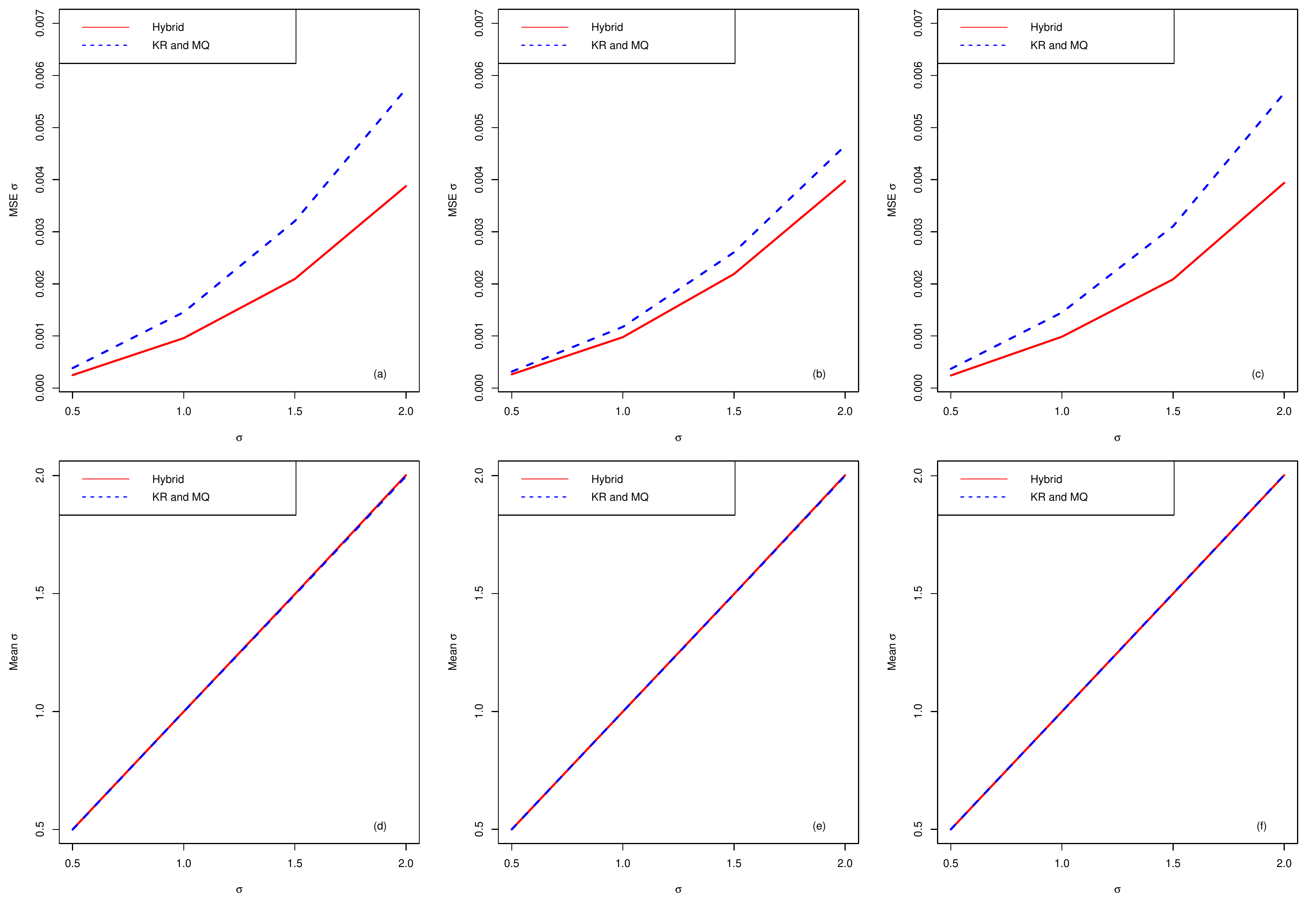}
\caption{Mean and MSE of $\sigma$}
\label{222}
\end{figure}

\begin{figure}
\includegraphics[scale=0.5, angle=0]{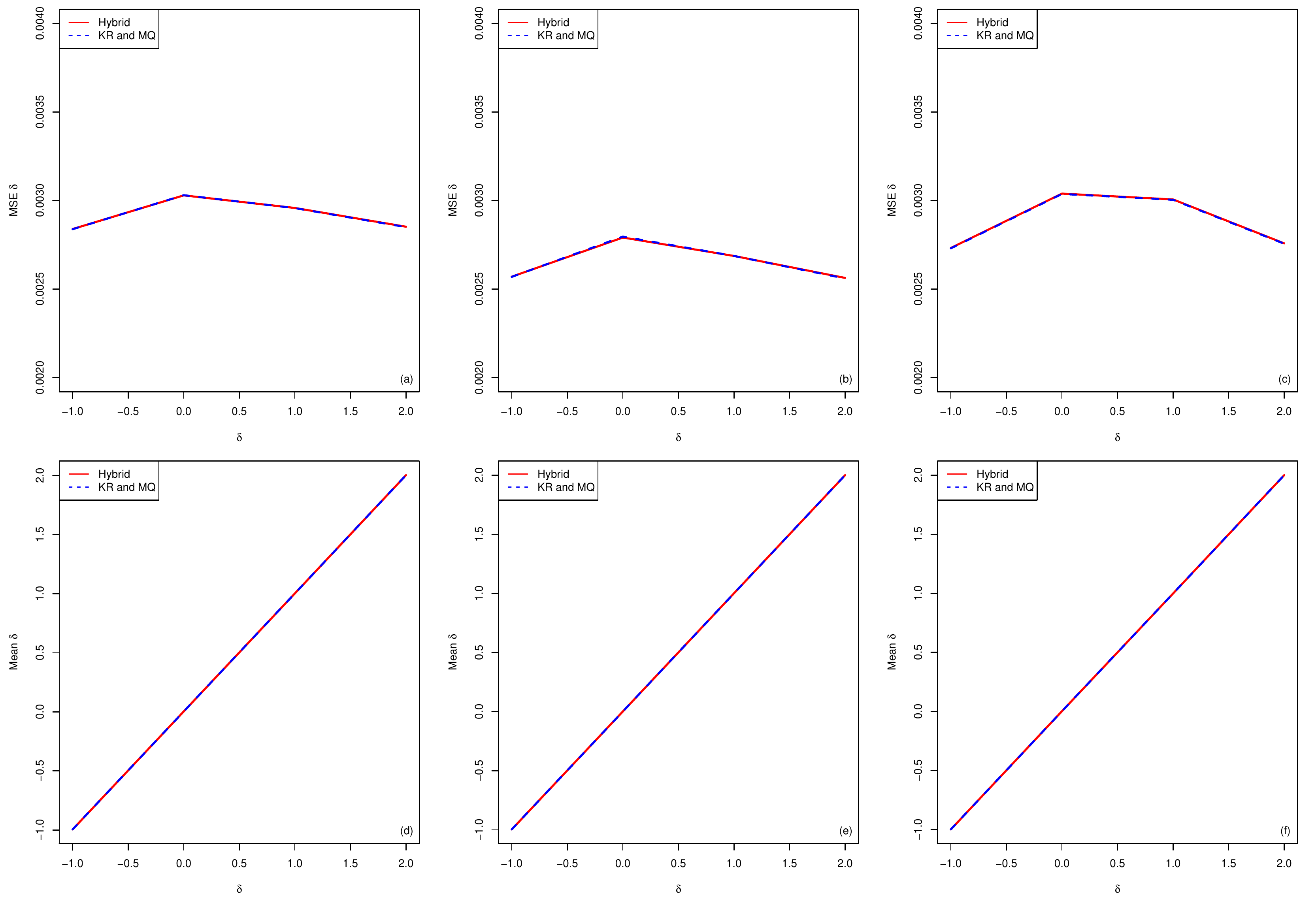}
\caption{Mean and MSE of $\delta$}
\label{333}
\end{figure}

Tables \ref{11}, \ref{22} and \ref{33} compare the MSE, mean and Sd, while Figures \ref{111}, \ref{222} and \ref{333} compare the MSE and mean of all the three methods employed for the estimation of the parameters $\alpha$, $\sigma$ and $\delta$. 

\noindent
For estimating $\alpha$, the parameters $\sigma$, $\delta$ and the sample size \textit{n} is fixed to 1, 0 and 1500 respectively. The parameter $\alpha$ is allowed to vary from 0.4 to 2 with a step size of 0.4. For obtaining the estimate of $\sigma$, fix $\alpha$, $\delta$ and \textit{n} to 1.3, 0 and 1500 respectively with $\sigma$ varying from 0.5 to 2 with a step of 0.5. Lastly, for the estimation of $\delta$, set $\alpha$, $\sigma$ and \textit{n} to 1.4, 1, 1500 respectively.

\noindent
From Table \ref{11} and Figure \ref{111}, it is quite evident that the MSE and Sd of the hybrid method, when estimating $\alpha$ is much lower than the KR and MQ method for $\beta \in \lbrace  -0.5, 0, 0.5 \rbrace$ thereby depicting the stability of our method. However only for $\alpha=2$, KR outperforms our method. The value of $\alpha$ estimated by the MQ method diverges greatly due to large MSE values specifically in the case of $\alpha=2$ and $\alpha < 0.6$ while the method of KR  slightly gives less accurate results for smaller values of $\alpha$.

\noindent
From Table \ref{22} and Figure \ref{222}, we observe that 
the MSE values increase as $\sigma$ increases, however, the values obtained via the hybrid method are comparatively lower than KR and MQ. Significant differences in the values can be seen when $\sigma \geq 1.$

\noindent
Table \ref{33} and Figure \ref{333} show that the means of $\delta$ estimated by the three methods are all very close to the true value. The MSE values obtained via hybrid method is almost at par with KR and MQ corresponding to $\beta=-0.5$, $\beta=0$ and $\beta=0.5.$ 

\noindent
Thus, through our simulations, we conclude that the hybrid method has the best estimation accuracy with low MSE values for the three parameters $\alpha, \sigma$ and $\delta$ followed by KR and then MQ.


\begin{table}[h]
\centering \footnotesize
\begin{tabular}{| *{11}{c |}} 
\hline
\textbf{Values} & \textbf{$\alpha$} & \multicolumn{3}{| c }{$\beta=-0.5$} & \multicolumn{3}{| c }{$\beta=0$}& \multicolumn{3}{| c |}{$\beta=0.5$}\\
\hline
& & \textbf{Hybrid } & \textbf{KR} & \textbf{MQ} & \textbf{Hybrid}     & \textbf{KR} &\textbf{MQ}& \textbf{Hybrid } & \textbf{KR} & \textbf{MQ} \\
 \hline
MSE& & 0.000426 &0.000611 &0.01 &0.000413 &0.000664 &0.01&0.000426&0.0006115&0.01\\
Mean & 0.4  & 0.398433     & 0.400449   & 0.5 &  0.398994     & 0.400687 &0.5 & 0.398434& 0.400449&0.5\\
Sd &   & 0.020602  &0.024724  & - & 0.020298  & 0.025774  &-& 0.020602&0.024724&-\\ 
 \hline
MSE  &  & 0.000943     & 0.001117       & 0.001771 & 0.0009571     & 0.001164    & 0.009581&0.000943&0.001171&0.001771\\
Mean  & 0.8  & 0.802503   &  0.804636    & 0.802713 & 0.801734     & 0.802408    & 0.795937&0.802503&0.804636&0.802713 \\
Sd &   & 0.030614  & 0.033909 & 0.041999 &0.030889 &0.034040 & 0.097803&0.030614&0.033909&0.041999\\
\hline    
 MSE  &  &  0.001357   & 0.001794 &0.002439 &  0.001455    &   0.001927   &  0.002734 &0.001257&0.001294&0.002439\\
 Mean  & 1.2  & 1.200390   & 1.201220   & 1.200063& 1.198824   & 1.198919 &1.198988&1.200390&1.201220&1.200063\\
 Sd &   & 0.036842 &0.042339  & 0.049394& 0.038138  & 0.043887  &  0.052278&0.036842&0.042339&0.049394\\ 
\hline
MSE  &  & 0.001978    &0.002260   & 0.003865 &   0.001848  &0.002206     &0.003411&0.001978&0.002260&0.003865 \\
Mean  & 1.6  &  1.604247   & 1.603363      & 1.604000 &  1.603294 &  1.602019&1.603023    & 1.604247&1.603363&1.604000\\
Sd &   &  0.044273 & 0.047423 &0.062041& 0.042868   & 0.046932 &0.058331&0.044273&0.047423&0.062041 \\
 \hline
 MSE  &  & 0.001744      &0.000044   & 1.242701 &   0.001726   &0.000048 &1.177734&0.001744&0.000044 &1.242701 \\
 Mean  & 2  &  1.996400   & 1.995918     & 1.134903&  1.993769 &  1.995718    &  1.175857&1.996400&1.995918&1.134903\\
Sd &   &  0.041609 & 0.005238 &0.703070& 0.041081   & 0.005458 &0.706060&0.041609&0.005238&0.703070 \\
 \hline
\end{tabular} 
\caption{Estimation of $\alpha$ for  $n=1500,~ \sigma =1,~ \delta = 0$}
\label{11}
\end{table}
\begin{table}[h]
\centering \footnotesize
\begin{tabular}{| *{8}{c |}} 
\hline
\textbf{Values} & \textbf{$\sigma$} & \multicolumn{2}{| c }{$\beta=-0.5$} & \multicolumn{2}{| c }{$\beta=0$}& \multicolumn{2}{| c |}{$\beta=0.5$}\\
\hline
   & & \textbf{Hybrid } & \textbf{KR} \textbf{and} \textbf{MQ} & \textbf{Hybrid}     & \textbf{KR} \textbf{and} \textbf{MQ}& \textbf{Hybrid } & \textbf{KR} \textbf{and} \textbf{MQ} \\
 \hline
MSE& & 0.000249 &0.000383 &0.000264 &0.000316  &0.000242 &0.000369\\
Mean & 0.5  & 0.499935     & 0.499184    & 0.499796 &  0.500027     & 0.499803       & 0.500395\\
Sd &   & 0.015783  &0.019572  & 0.016262 & 0.017792  & 0.015581  & 0.019227\\ 
 \hline
MSE  &  & 0.000959     & 0.001457       & 0.000977 & 0.001175     & 0.000986    & 0.001451\\
Mean  & 1  & 0.999559   &  0.999462    & 0.999417 & 0.999345      & 0.999984    & 0.999723 \\
Sd &   & 0.030966  & 0.038168 & 0.031265 &0.034276 &0.031415  & 0.038097\\
\hline    
 MSE  &  &  0.002094    & 0.003205       &0.002189 &  0.002606    &   0.002088   &  0.003106  \\
 Mean  & 1.5  & 1.497367   & 1.494755   & 1.498057& 1.497125    & 1.499892     &1.500856\\
 Sd &   & 0.045692 &0.056376  & 0.046748& 0.050973  & 0.045700  &  0.055725\\ 
\hline
MSE  &  & 0.003877      &0.005743    & 0.003976 &   0.004651   &0.003938     &0.005662 \\
Mean  & 2  &  2.001812    & 1.997251      & 2.002375 &  1.999884 &  2.002557    &  2.003091\\
Sd &   &  0.062241 & 0.075739 &0.063016& 0.068202    & 0.062704 &0.075188 \\
 \hline
\end{tabular} 
\caption{Estimation of $\sigma$ for  $n=1500,~ \alpha =1.3,~ \delta = 0$}
\label{22}
\end{table}

\begin{table}[h!]
\centering \footnotesize
\begin{tabular}{| *{8}{c |}} 
\hline
\textbf{Values} & \textbf{$\delta$} & \multicolumn{2}{| c }{$\beta=-0.5$} & \multicolumn{2}{| c }{$\beta=0$}& \multicolumn{2}{| c |}{$\beta=0.5$}\\
\hline
   & & \textbf{Hybrid } & \textbf{KR}         \textbf{and} \textbf{MQ} & \textbf{Hybrid}     & \textbf{KR}  \textbf{and} \textbf{MQ}& \textbf{Hybrid } & \textbf{KR} \textbf{and}      \textbf{MQ} \\
 \hline   
MSE& & 0.002839 &0.002838 &0.002570 &0.002569  &0.002732 &0.002730\\
Mean & -1 & -0.99527     & -0.99529   & -0.996797 & -0.996793     & -0.999462       & -0.999458\\
Sd &   & 0.053082  &0.053072 & 0.050594 & 0.050586  & 0.015581  & 0.019227\\ 
 \hline
MSE  &  & 0.003029    & 0.003030       & 0.002791 & 0.002796     & 0.003039    & 0.003037\\
Mean  & 0  & 0.003340  &  0.003339    & 0.001437 & 0.001478      & 0.001555   & 0.001581 \\
Sd &   & 0.054943 & 0.054950 & 0.052818 &0.052856 &0.055108  & 0.055092\\
\hline    
 MSE  &  &  0.002958  & 0.002957       &0.002687 &  0.002687    &   0.003006   &  0.003004  \\
 Mean  & 1 & 1.002198   & 1.002197   & 1.001027& 1.001055    & 0.998137     &0.998154\\
 Sd &   & 0.054343 &0.054340  & 0.051833& 0.051826  & 0.054799  &  0.054786\\ 
\hline
MSE  &  & 0.002852     &0.002850   & 0.002563 &   0.002560   &0.002758 &0.002756\\
Mean  & 2  &  2.003242   & 2.003252      & 2.000933 &  2.000936 &  2.000278    &  2.000271\\
Sd &   &  0.053306 & 0.053295 &0.050624& 0.050588    & 0.052517 &0.052497 \\
 \hline
\end{tabular} 
\caption{Estimation of $\delta$ for  $n=1500,~ \alpha =1.4,~ \sigma = 1$}
\label{33}
\end{table}

\clearpage
\subsection{Examples of the Estimation of $\Gamma$ in 1-$d$, 2-$d$ and the 3-$d$ Case}
The data is simulated using the Modarres and Nolan \cite{mod} procedure. In the univariate and trivarate case, we observe that the estimates, obtained after the implementation of our method, are accurate with low RMSE (root mean squared error) as shown in Tables \ref{40}, \ref{70} and \ref{80}.  For the bivariate case, the R package ``alphastable" is used to perform all the simulations. The estimates obtained in this case, are compared with the method proposed by Mohammadi \textsl{et al.}\cite{moh} in terms of the mean and RMSE (root mean squared error). Tables \ref{60} and \ref{50} showcase that our method outperforms Mohammadi \textsl{et al.} \cite{moh} method. The new method is not compared to the method of Teimouri \textsl{et al.} \cite{tei} due to high time complexity.  \\
\begin{example}
For $d=1$, the univariate data is simulated from $\alpha$=1.6, sample size $(n)=1200$, $L=2$, $\gamma_{l}=1/2$, $\mathbf{s}_{l}=(-1)^{l}$, $\mathbf{t}_{l}=(-1)^{l+1}$ for $l=1,~2$, location vector $\boldsymbol \delta=(0,0)$ with $\psi_{\alpha}(u)$ and $\hat{\mathbf{I}}$ defined in (\ref{8}) and 100 iterations.
\end{example}
\begin{table}[h]
\centering 
\begin{tabular}{| *{3}{c |}} 
\hline
\textbf{Values}  & \multicolumn{2}{| c | }{\textbf{Proposed Method}}  \\
\hline
 & \textbf{mean } & \textbf{RMSE} \\
 \hline
$\alpha$ & 1.6000 & 0.047 \\
 \hline
 $\gamma_{1}$ & 0.5033 & 0.029 \\
 \hline
 $\gamma_{2}$ & 0.4963 & 0.032\\
 \hline
\end{tabular} 
\caption{Estimation of $\Gamma$ for  $n=1200,~ \alpha =1.6,~ \delta=(0,0),~\gamma_{l}=1/2$}
\label{40}
\end{table}
\begin{example}
For $d=2$, we considered the bivariate data simulated from $\alpha$=1.3, sample size $(n)=1300$, $L=4$, $\gamma_{l}=1/4$, $\mathbf{s}_{l}=\mathbf{t}_{l}=\big(\cos(2\pi(l-1)/L,\sin(2\pi(l-1)/L\big)$ for $l=1,\cdots, 4$, location vector $\boldsymbol \delta=(0,0)$ with $\psi_{\alpha}(u)$ and $\hat{\mathbf{I}}$ defined as in the method suggested by Nolan when L is even and 500 iterations.
\end{example}
\begin{table}[h]
\centering 
\begin{tabular}{| *{5}{c |}} 
\hline
\textbf{Values}  & \multicolumn{2}{| c }{\textbf{Proposed Method}} & \multicolumn{2}{| c |}{\textbf{Mohammadi \textsl{et al.}}} \\
\hline
 & \textbf{mean } & \textbf{RMSE} & \textbf{mean}  & \textbf{RMSE}\\
 \hline
$\alpha$ & 1.3017 & 0.0302 & 1.3054  & 0.0604\\
 \hline
 $\gamma_{1}$ & 0.2500 & 0.0147 & 0.2504   &0.0146 \\
 \hline
 $\gamma_{2}$ & 0.2495 & 0.0145 & 0.2495   &0.0155 \\
 \hline
 $\gamma_{3}$ & 0.2506 & 0.0142 & 0.2502   &0.0149 \\
 \hline
 $\gamma_{4}$ & 0.2501 & 0.0144 & 0.2490   &0.0148 \\
 \hline
 \end{tabular} 
\caption{Estimation of $\Gamma$ for  $n=1300,~ \alpha =1.3,~ \delta=(0,0),~\gamma_{l}=1/4$}
\label{60}
\end{table}
 \begin{example}
For $d=2$, we considered the bivariate data simulated from $\alpha$=1.5, sample size $(n)=1300$, $L=5$, $\gamma_{l}=1/5$, $\mathbf{s}_{l}=\mathbf{t}_{l}=\big(\cos(2\pi(l-1)/L,\sin(2\pi(l-1)/L\big)$ for $l=1,\cdots, 5$, location vector $\boldsymbol \delta=(0,0)$ with $\psi_{\alpha}(u)$ and $\hat{\mathbf{I}}$ defined as in the method suggested by Nolan and 500 iterations.
\end{example}
\begin{table}[h]
\centering 
\begin{tabular}{| *{5}{c |}} 
\hline
\textbf{Values}  & \multicolumn{2}{| c }{\textbf{Proposed Method}} & \multicolumn{2}{| c |}{\textbf{Mohammadi \textsl{et al.}}} \\
\hline
 & \textbf{mean } & \textbf{RMSE} & \textbf{mean}  & \textbf{RMSE}\\
 \hline
$\alpha$ & 1.500 & 0.0387 & 1.5032  & 0.0753\\
 \hline
 $\gamma_{1}$ & 0.2115 & 0.0218 & 0.2130   &0.0229 \\
 \hline
 $\gamma_{2}$ & 0.2041 & 0.0185 & 0.2045   &0.0200 \\
 \hline
 $\gamma_{3}$ & 0.1915 & 0.0207 & 0.1909   &0.0222 \\
 \hline
 $\gamma_{4}$ & 0.1892 & 0.0209 & 0.1889   &0.0231 \\
 \hline
 $\gamma_{5}$ & 0.2046 & 0.0182 & 0.2048   &0.0195 \\
 \hline
 \end{tabular} 
\caption{Estimation of $\Gamma$ for  $n=1300,~ \alpha =1.5,~ \delta=(0,0),~\gamma_{l}=1/5$}
\label{50}
\end{table}
\newpage
\begin{example}
For $d=3$, we simulated from $\alpha$=1.7, sample size $(n)=1400$, $L=3$, $\gamma_{l}=1/3$, $\mathbf{s}_{l}=\mathbf{t}_{l}=\Big(\sin(\pi/l)\cos(2\pi(l-1)/L),~\sin(\pi/l)\sin(2\pi(l-1)/L),~\cos(\pi/l)\Big),~l=1,\cdots, 3$, location vector $\boldsymbol \delta=(0,0)$ with $\psi_{\alpha}(u)$ defined in (\ref{811}) and $\hat{\mathbf{I}}=\big(\Re(\hat{I}(t_{1}),~\Re(\hat{I}(t_{2}),\cdots,\Re(\hat{I}(t_{L})\big)$
\end{example}
\begin{table}[h]
\centering 
\begin{tabular}{| *{5}{c |}} 
\hline
\textbf{Values}  & \multicolumn{2}{| c |}{\textbf{Proposed Method}} \\
\hline
 & \textbf{mean } & \textbf{RMSE} \\
 \hline
$\alpha$ & 1.6989 & 0.0361 \\
 \hline
 $\gamma_{1}$ &  0.3322 & 0.0180  \\
 \hline
 $\gamma_{2}$ &  0.3335 & 0.0179  \\
 \hline
 $\gamma_{3}$ & 0.3361 &  0.0203 \\
 \hline
 \end{tabular} 
\caption{Estimation of $\Gamma$ for  $n=1400,~ \alpha =1.7,~ \delta=(0,0),~\gamma_{l}=1/3$}
\label{70}
\end{table}

\begin{example}
For $d=3$, we simulated from $\alpha$=1.8, sample size $(n)=1300$, $L=4$, $\gamma_{l}=1/4$, $\mathbf{s}_{l}=\mathbf{t}_{l}=(\sin(\pi/l)\cos(2\pi(l-1)/L),~\sin(\pi/l)\sin(2\pi(l-1)/L),~\cos(\pi/l)),~l=1,\cdots, 4$, location vector $\boldsymbol \delta=(0,0)$ with $\psi_{\alpha}(u)$ defined in (\ref{811}) and $\hat{\mathbf{I}}=(\Re(\hat{I}(t_{1}),~\Re(\hat{I}(t_{2}),\cdots,\Re(\hat{I}(t_{L}))$
\end{example}
\begin{table}[h]
\centering 
\begin{tabular}{| *{5}{c |}} 
\hline
\textbf{Values}  & \multicolumn{2}{| c |}{\textbf{Proposed Method}} \\
\hline
 & \textbf{mean } & \textbf{RMSE} \\
 \hline
$\alpha$ & 1.800 & 0.0326 \\
 \hline
 $\gamma_{1}$ &  0.2504 & 0.0216  \\
 \hline
 $\gamma_{2}$ &  0.2501 & 0.0227  \\
 \hline
 $\gamma_{3}$ & 0.2501 &   0.0140 \\
 \hline
 $\gamma_{4}$ & 0.2491 &  0.0346 \\
 \hline
 \end{tabular} 
\caption{Estimation of $\Gamma$ for  $n=1300,~ \alpha =1.8,~ \delta=(0,0),~\gamma_{l}=1/4$}
\label{80}
\end{table}
\section{Applications to Financial Data}
\label{sec5}
In this section, we apply our above introduced method to real life data. We illustrate two examples based on univariate and strictly bivariate stable distribution.
\subsection*{International Business Machines Corporation (IBM)}
\begin{figure}[h]
\includegraphics[scale=0.6]{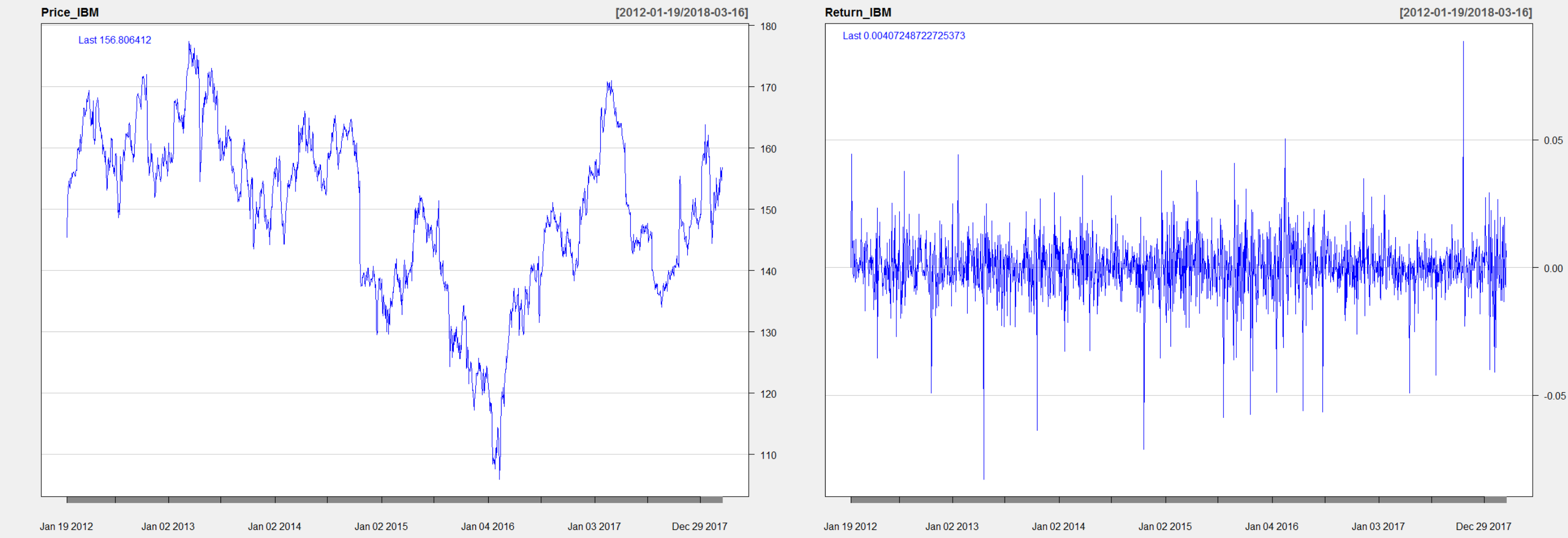}
\caption{Price and Return of IBM}
\label{25}
\end{figure}

\begin{figure}[h]
\includegraphics[scale=0.6]{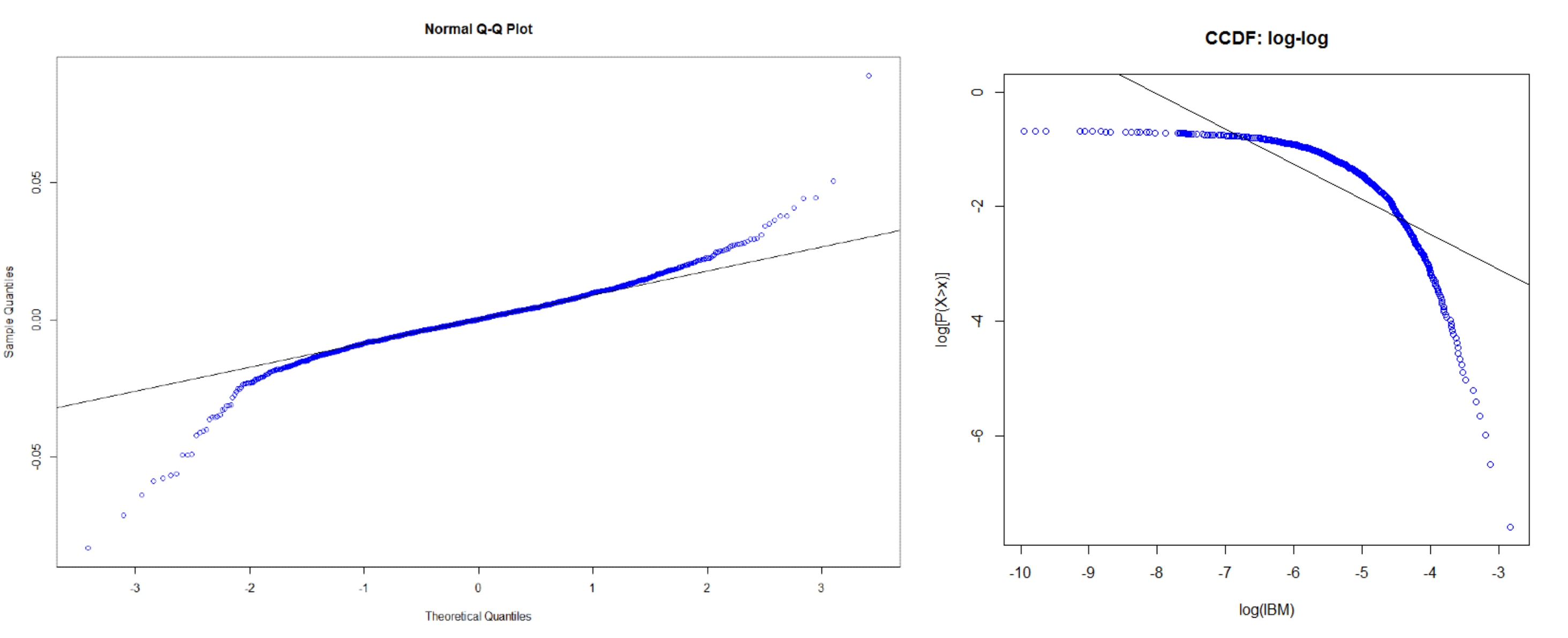}
\caption{QQ-plot and CCDF plot}
\label{26}
\end{figure}

\begin{figure}
\includegraphics[width=7in, height=5in]{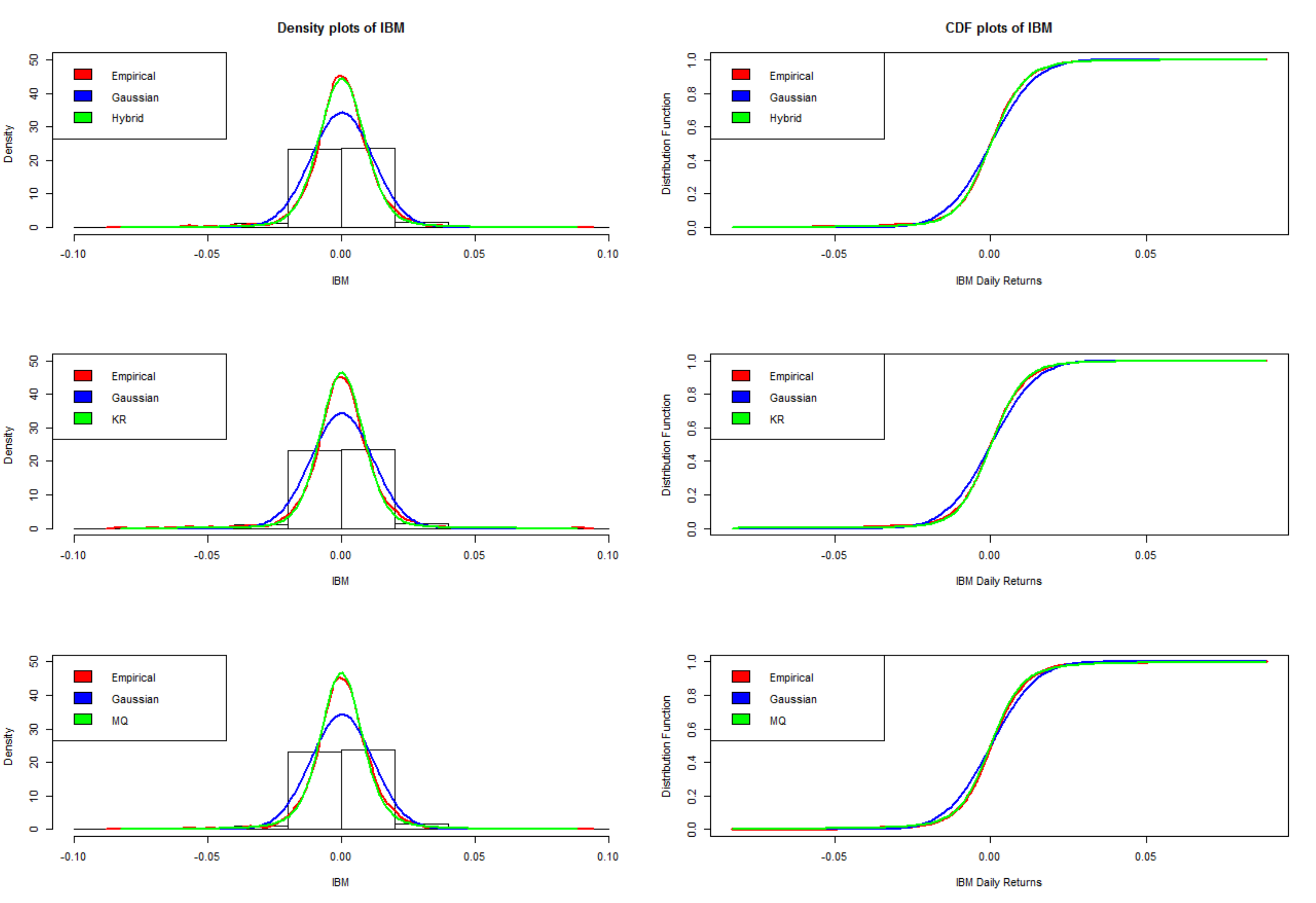}[h]
\caption{Density and CDF plots of IBM}
\label{6}
\end{figure}

The data has been obtained from Yahoo Finance for the datasets dealing with the prices of stock for our empirical analysis. The first dataset that we have considered is, International Business Machines Corporation (IBM) from New York Stock Exchange for the period January 19, 2012-March 19, 2018 comprising of 1550 daily returns value of the adjusted closing price. The price and return of IBM adjusted closing price are depicted in Figure \ref{25}.\\
\\
In order to ensure that the given dataset can be modelled by a heavy tailed distribution specifically a non-Gaussian stable distribution, some plots and normality tests have been carried out. The $p$-values obtained via Anderson-Darling and Shapiro-Wilk test were far lesser than 0.05 thereby disabling us to accept the null hypothesis on the support of normality.  From the QQ-normal plot in Figure \ref{26}, it is again evident that the data is not normally distributed as the points in the plot do not lie on a straight diagonal line. Also the plot of the empirical cumulative distribution function on a log-log scale, in Figure \ref{26}, shows that the dataset cannot even be modelled by a power law distribution which is another type of heavy-tailed distribution.\\
\\
Next, we check which model whether Gaussian or non-Gaussian stable model, namely, Hybrid, KR, MQ and Maximum Likelihood(ML) gives a better fit to the observed returns of IBM. The estimates obtained from each  model are given in Table \ref{4}. The performance of each of the models is graphically assessed through the density and cumulative distribution function plots as seen in Figure \ref{6}. The theoretical assessment is done via the Kolmogorov-Smirnov(K-S) goodness of fit test. The K-S statistic measures the distance between the empirical cumulative distribution function (ECDF) of the sample data and the cumulative distribution function (CDF) of the reference distribution. It is defined as 
$$\rm D =\sup_{\textit{x}\in \mathbb{R}}\mid\mid \textit{F}(\textit{x};\alpha,\beta,\sigma,\delta)-\hat{\textit{F}}(\textit{x};\alpha,\beta,\sigma,\delta)\mid\mid$$
where sup is the supremum, $F$ and $\hat{F}$ denote the ECDF and CDF computed from the estimated probability density function. The D values and the $p$-values obtained using different models are given Table \ref{5}.
\begin{table}[h]
\centering
\begin{tabular}{| c | c | c | c | c | c |}
 \hline
 & Method & $\alpha$ & $\beta$ & $\sigma$ & $\delta$ \\
\hline
 &MQ  & 1.6050 & 0.0450   & 0.0061 & 0.000073 \\
stable fit & KR  &  1.7518 & 0.0751& 0.0061& 0.000073\\  
& Hybrid & 1.7463  &0.0751 & 0.0063  &0.000071   \\
&ML&1.7518&0.0751&0.0064&0.000194\\
 \hline
\end{tabular}
 \caption{stable fit to 1550 IBM daily returns of the adjusted closing price obtained using different models}
\label{4}
 \end{table}
\begin{table}[h]
\centering
\begin{tabular}{| c | c | c | c |}
\hline
& Method & D & $p$-value\\
\hline
& Hybrid & 0.018538& 0.6612   \\
& MQ &  0.018603 & 0.6568\\
K-S test& KR & 0.019432  &0.6019   \\
&ML& 0.020233&0.5497\\
&Gaussian&0.074748&$6.01\times 10^{-8}$ \\
\hline
\end{tabular}
\caption{K-S goodness of fit test to 1550 IBM daily returns of the adjusted closing price obtained using different models}
\label{5}
\end{table}\\
From Figure \ref{6}, we observe that the density of the normal distribution is too low near the middle, high in the midrange and quite low on the tails. While on the other hand, it is interesting to observe how all the stable models specifically the hybrid model approximate the data well over almost the whole range. Also from Table \ref{5} the K-S goodness of fit test reveals that D is the smallest for the hybrid. 
\newpage
\subsection*{Tata Consultancy Services Limited (TCS.NS) and National Thermal Power Corporation Limited (NTPC.NS)}
\begin{figure}[h]
\centering
\includegraphics[width=7in, height=3in]{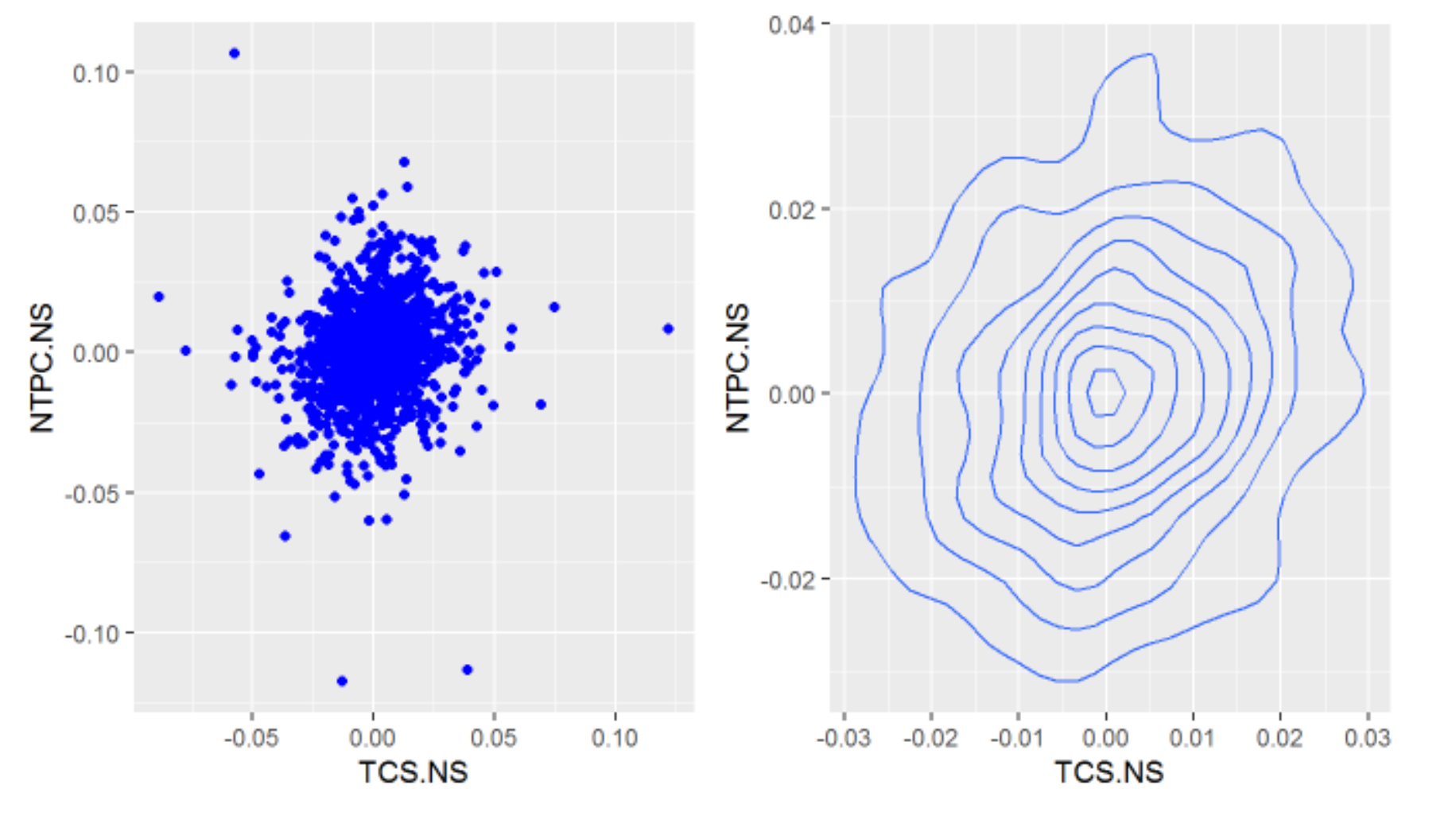}
\caption{Scatter and Contour plot of TCS.NS and NTPC.NS}
\label{77}
\end{figure}

The daily returns of 1475 adjusted closing prices for the two components of Nifty, namely NTPC.NS and TCS.NS is obtained from Yahoo Finance for the time period, January 3, 2011 to December 31, 2016. Figure \ref{77} shows the scatter and the contour plots of their returns respectively and reveals that the data is heavily skewed downwards and several points are away from the origin. Also from the contour plot it is clear that the distribution of the data is neither normal nor elliptical stable.
Thus, multivariate stable distributions are more appropriate than multivariate normal or elliptical distributions. The theoretical assessment of whether the data is multivariate normal is done using the MVN package in R and the test used is Henze-Zirkler test. The test reveals that the data is not multivariate normal. The details of this test on the bivariate data are shown in Table \ref{88}.
\begin{table}[h]\footnotesize
\begin{tabular}{| c | c | c | c | c | c | c | c | c | c |}
\hline
Test & Statistic & $p$-value & Normality & Mean & Standard Deviation & Median & Skew & Kurtosis \\
\hline
 Henz-Zirkler & 7.2338& 0 & NO & & & & &   \\
 \hline
 Shapiro-Wilk (TCS.NS) &  0.9681 & $<0.001$ & NO & 0.00074 & 0.01616 & 0.00035 & 0.14523 & 3.9688\\
\hline
 Shapiro-Wilk (NTPC.NS)& 0.9676 & $<0.001$ & NO & 0.00012 & 0.01692 & 0 &-0.19950 & 4.2812 \\
\hline
\end{tabular}
\caption{Results obtained using the mvn package in R}
\label{88}
\end{table}

\noindent
We now model the pair $\mathbf{X}=$( TCS.NS, NTPC.NS) using a bivariate stable distribution with $L=12$ points of masses for spectral measure given by $\mathbf{t}_{l}=\mathbf{s}_{l}=\big(\cos(2\pi(l-1)/L,\sin(2\pi(l-1)/L\big)\in S^{2}$, $l=1, 2, \cdots, L$. The estimates of the location vector $\hat{\boldsymbol\delta}$ obtained using our proposed 
hybrid method are $\hat{\boldsymbol\delta}=(0.0003077732, -0.0001562171)$ while $\hat{\alpha}=1.85591$, obtained after taking the mean of $\hat{\alpha}_{1}=1.841935$ and $\hat{\alpha}_{2}=1.87007$. To fit a strictly stable distribution transform $\mathbf{X}$ to $\mathbf{X}-\hat{\boldsymbol\delta}$. The results of the estimates of the spectral measure after fitting a strictly stable distribution to the shifted data is shown in Table \ref{99} and is compared with other well-known methods such as Mohammadi \textsl{et al.} (M) \cite{moh} and Teimouri \textsl{et al.} (T) \cite{tei}. We observe that the values of the estimated masses obtained via our method are very much closer to the other two methods.
\begin{table}[h]\footnotesize
\centering
\begin{tabular}{| c | c | c | c | c | c | c | c | c | c | c | c | c | c |}
\hline
 Methods & $\alpha$ & $\gamma_{1}$ & $\gamma_{2}$ & $\gamma_{3}$  & $\gamma_{4}$  & $\gamma_{5}$  & $\gamma_{6}$  &$\gamma_{7}$  & $\gamma_{8}$ & $\gamma_{9}$ & $\gamma_{10}$ & $\gamma_{11}$ & $\gamma_{12}$  \\
\hline
M& 1.8349  & 0 & 0  & 0  & 0 & 0  &0  & 0.00017 & 0 & 0.00019 & 0.00002 & 0& 0  \\
\hline
T& 1.6802  & 0 & 0  & 0  & 0 & 0  &0  & 0.00020 & 0 & 0.00025 & 0 & 0& 0  \\
\hline
Proposed Method & 1.8559  & 0 & 0  & 0  & 0 & 0  &0  & 0 & 0 & 0.00020 & 0 & 0.00012& 0.00005 \\
\hline
\end{tabular}
\caption{Estimates of the spectral measure obtained using different method for \newline a strictly stable bivariate data}
\label{99}
\end{table}

\section{Concluding Remarks}
To conclude, we make the following observations in relation to our proposed method.
\begin{itemize}
\item[1.] Our proposed hybrid method (univariate) is non-iterative in nature as no further improvement in the MSE's of the estimators resulted after the first iteration. Also, as in the case of Kogon-Williams method, the regressions are performed only once using ordinary least squares as opposed to the numerous iterations for Koutrouvelis' method where regressions are performed using generalized least squares. Thus, in terms of computational efficiency, our method is as good as the method of Kogon-Williams.

\item[2.] Koutrouvelis makes use of look-up tables in order to perform regressions while Kogon-Williams eliminated the need to use look-up tables for estimation. Our proposed hybrid method shows that if we retain the look-up table and modify it in order to obtain the regression estimates using ordinary least squares, the performance of the estimates improves significantly in comparison to the estimates obtained using Kogon-Williams method in terms of accuracy and low MSE.

\item[3.] The method of moments by Press is said to yield poor estimates, however, in our proposed hybrid method we have used this method to get the estimate of $\delta$ by suitably choosing two points at which the sample characteristic function is evaluated. Thus, by using the Press' method we obtained a very good estimate of $\delta$.

\item[4.] For the multivariate case, we make use of our proposed hybrid method to obtain the estimators $\hat{\alpha}$ and the shift vector $\hat{\boldsymbol \delta}$. The discretized estimator $\hat{\boldsymbol \gamma}$ of the spectral measure $\Gamma$ is obtained via the empirical characteristic method suggested by Nolan \textsl{et al.} \cite{pan} with slight modifications. Thus, in terms of computational efficiency and  accuracy, the new method outperforms the method of Mohammadi \textsl{et al.} (M) \cite{moh} and Teimouri \textsl{et al.} (T) \cite{tei}.

\item[5.] Finally, we give two applications of our proposed method using financial data, where, the distribution of the datasets considered is stable. For the univariate data, K-S goodness of fit test shows that our method best fits the data in comparison to the other methods. Though the maximum likelihood method of estimation is said to give the most accurate estimate, however, it is computationally, the slowest when applied to the two financial data. For the bivariate data, we observe that the values of the estimated masses obtained via our method are very close to the values obtained through the method of Mohammadi \textsl{et al.} (M) \cite{moh} and Teimouri \textsl{et al.} (T) \cite{tei}.
\end{itemize}

\end{document}